\documentclass[graybox]{svmult}

\usepackage{mathptmx}       
\usepackage{helvet}         
\usepackage{courier}        
\usepackage{makeidx}         
\usepackage{graphicx}        
\usepackage{multicol}        

\makeindex             

\begin{document}

\title*{Nature and nurture of early-type dwarf galaxies in low density environments}
\author{Ruth Gr\"utzbauch, F. Annibali, R. Rampazzo, A. Bressan, W. W. Zeilinger}
\institute{Ruth Gr\"utzbauch\at University of Nottingham, UK, \email{ruth.grutzbauch@nottingham.ac.uk}\\
F. Annibali, R. Rampazzo, A. Bressan \at INAF – Osservatorio Astronomico di Padova, Italy\\
W. W. Zeilinger\at Department of Astronomy, Univ. of Vienna, Austria
}
%
%
\maketitle

\vskip-1.2truein

\abstract{We study stellar population parameters of a sample of 13 dwarf galaxies located in poor groups of galaxies using high resolution spectra observed with VIMOS at the ESO-VLT \cite{Gru09}. LICK-indices were compared with Simple Stellar Population models to derive ages, metallicities and [$\alpha$/Fe]-ratios. 
Comparing the dwarfs with a sample of giant ETGs residing in comparable environments we find that the dwarfs are on average younger, less metal-rich, and less enhanced in alpha-elements than giants. Age, Z, and [$\alpha$/Fe] ratios are found to correlate both with velocity dispersion and with morphology. We also find possible evidence that low density environment (LDE) dwarfs experienced more prolonged star formation histories than Coma dwarfs, however, larger samples are needed to draw firm conclusions.
}

\section*{Age, Z, and [$\alpha$/Fe] as a function of velocity dispersion, morphology and environment}
\label{sec:2}

\begin{figure*}[t]
\begin{center}
\includegraphics[width=0.32\textwidth]{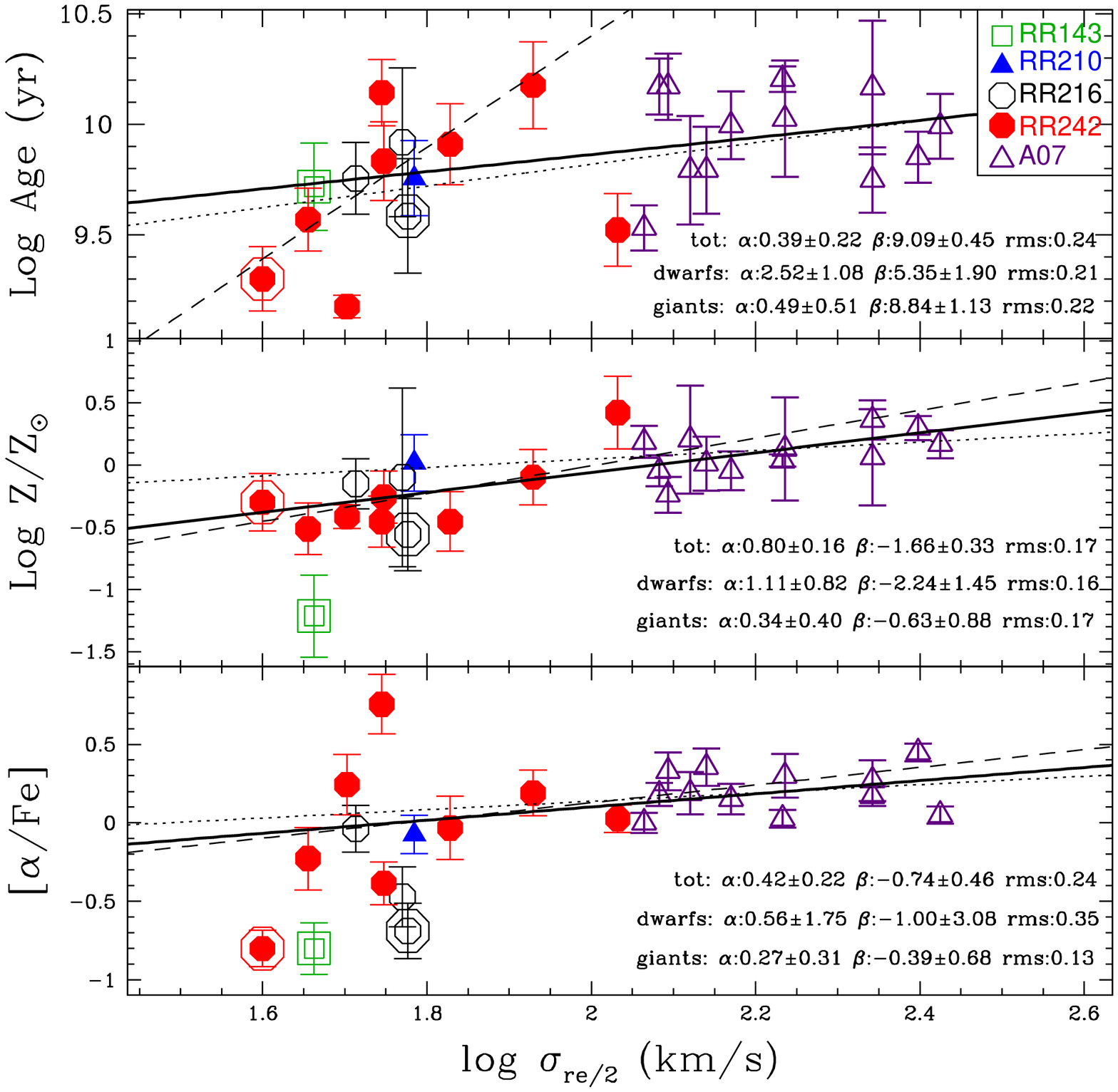}
\includegraphics[width=0.32\textwidth]{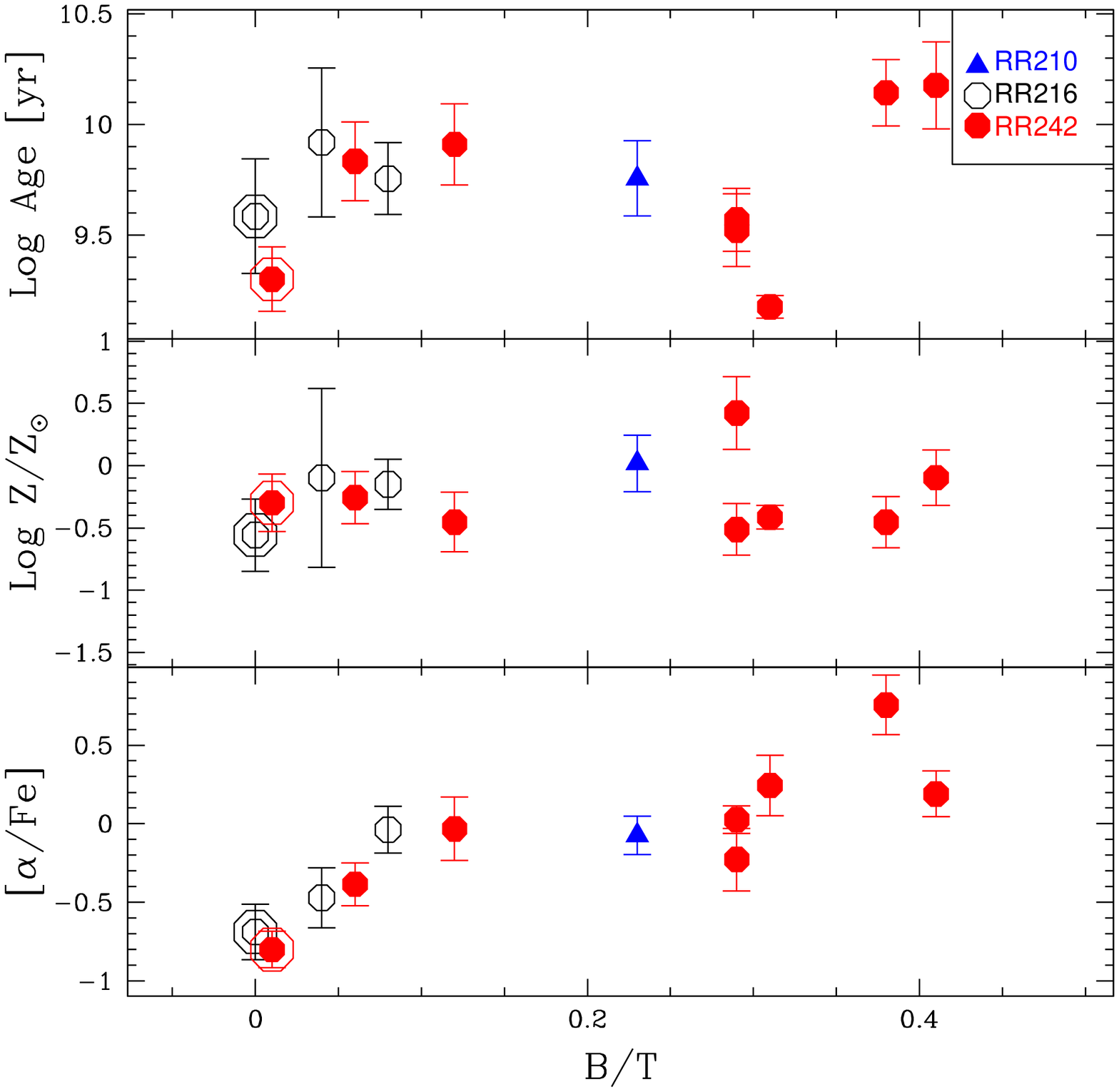}
\includegraphics[width=0.32\textwidth]{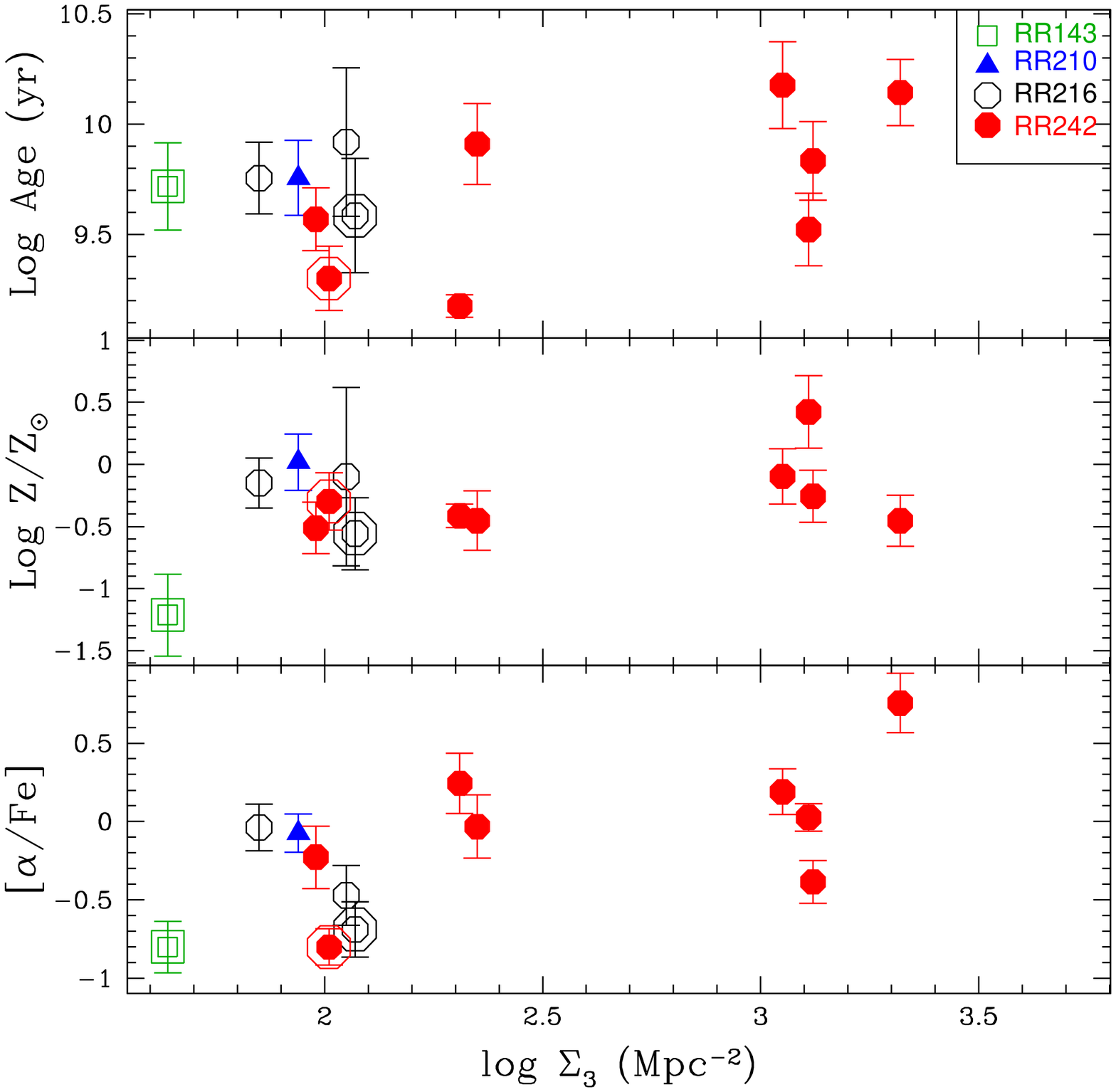}
\caption{Age, metallicity and [$\alpha$/Fe] as a function of velocity dispersion $\sigma$ (left) bulge fraction B/T (mid) and local density $\Sigma_3$ (right). Violet triangles are massive ETGs from \cite{Ann07}. \label{fig1}}
\end{center}
\end{figure*}

We confirm that the correlations between age, metallicity (Z), [$\alpha$/Fe] and velocity dispersion ($\sigma$) found for giant galaxies \cite{Ann07} extend towards the low-mass regime of our dwarf sample, albeit with a larger scatter (Figure~\ref{fig1}, left panel). The relatively tight Z-$\sigma$ relation, however, suggests that the scatter in age and [$\alpha$/Fe] is intrinsic, reflecting the variety of star formation histories in dwarf galaxies.

We find a strong correlation between the bulge fraction (B/T) and the [$\alpha$/Fe]-ratio, such that galaxies with a stronger bulge have shorter SF timescales (Figure~\ref{fig1}, mid panel). The same correlation is found for the Sersic index $n$. The presence of a morphology-[$\alpha$/Fe] relation seems in contradiction to the possible evolution along the Hubble sequence from low to high B/T galaxies. In other words, systems similar to present-day dwarf galaxies could not be the building blocks of today's giant ellipticals.

Local densities are measured for each galaxy based on the distance to the 3$^{rd}$ nearest neighbour. We find a weak correlation between local density ($\Sigma_3$) and [$\alpha$/Fe] for the dwarf sample, whereas age and metallicity are not clearly correlated with $\Sigma_3$ (Figure~\ref{fig1}, right panel). Note however, that $\Sigma_3$ is a projected density and projection effects can considerably distort the intrinsic 3D galaxy density.

\begin{figure}[t]
\begin{center}
\includegraphics[width=\textwidth]{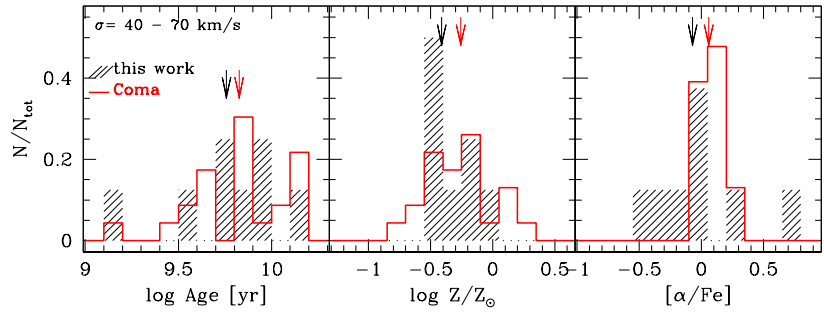}
\caption{Distribution of age, metallicity and [$\alpha$/Fe] in poor groups (shaded) compared to Coma (red). \label{fig2}}
\end{center}
\end{figure}

Finally, we compare our results to red, passive dwarf galaxies in the Coma Cluster studied by \cite{Smi09} (see Figure~\ref{fig2}). 
Statistical tests indicate that the only significant difference between our sample and the Coma sample is in the distribution of [$\alpha$/Fe], with LDE dwarfs having experienced a more prolonged star formation than cluster dwarfs.
By contrast, no strong difference in the [$\alpha$/Fe] ratios is observed for giant ETGs in the field and in the cluster.


\end{document}